\documentclass{article}
\usepackage[preprint, nonatbib]{nips}

\usepackage{amsmath,amssymb,amsfonts}
\usepackage{algorithmic}

\usepackage{graphicx}
\usepackage{epstopdf}

\usepackage{textcomp}
\usepackage{xcolor}
\usepackage{mathtools}
\usepackage{comment}
\usepackage{caption}
\usepackage{subcaption}
\usepackage{tabularx,booktabs}
\usepackage{slashbox}
\usepackage{booktabs,caption}
\usepackage[flushleft]{threeparttable}
\usepackage{multirow}

\usepackage{amsmath}
\usepackage{color}
\usepackage{comment}

\usepackage{mathtools}

\usepackage{overpic}

\usepackage{tabularx}

\newcolumntype{C}[1]{>{\centering\arraybackslash}p{#1}}

\author{%
  Tian Guo\thanks{Correspondence to \texttt{tig@ram-ai.com}}
  \And
  Nicolas Jamet \\
  \And
  Valentin Betrix \\
  \And
  Louis-Alexandre Piquet \\
  \And
  Emmanuel Hauptmann \\
  \\ 
  Systematic Equity Research \\ 
  RAM Active Investments \\
}

\begin{document}

\title{ESG2Risk: A Deep Learning Framework from ESG News to Stock Volatility Prediction}

\maketitle

\begin{abstract}
Incorporating environmental, social, and governance (ESG) considerations into systematic investments has drawn numerous attention recently. 
In this paper, we focus on the ESG events in financial news flow and exploring the predictive power of ESG related financial news on stock volatility.
In particular, we develop a pipeline of ESG news extraction, news representations, and Bayesian inference of deep learning models.
Experimental evaluation on real data and different markets demonstrates the superior predicting performance as well as the relation of high volatility prediction to stocks with potential high risk and low return.
It also shows the prospect of the proposed pipeline as a flexible predicting framework for various textual data and target variables.
\end{abstract}

\section{Introduction}

The widely adopted perspective for judging the sustainability of equity investments is along three pillars, E S G, which stand for \textbf{E}nvironmental, \textbf{S}ocial and \textbf{G}overnance.
In this paper we propose a novel approach of volatility forecasting based on ESG newsflow, an original integration of ESG into the investment process.
Recently, integrating sustainability into investment strategies are receiving exponentially increasing attention in finance \cite{de2015benefits, nagy2016can}. 
Environmental metrics cover all aspects of the firm's interaction with the environment, such as its CO2 emissions, its approach to the climate change transition or its broad strategy in the use of natural resources.
The Social dimension encompasses all standards set by companies as they build relationships with employees, suppliers and the communities in which they operate (labor conditions, equality, fairness to suppliers) while Governance would cover leadership elements like executive compensation, diversity of the board and controversies.
These ESG inputs are vital to assess the sustainability and the relevant risks of an investment position \cite{sassen2016impact}.
Conventionally, ESG related factors are formatted as structured data to facilitate the integration of ESG aspects into quantitative models and the building of expertise in systematic ESG investing.

In our research, we study the predictive power of ESG news on volatility. 
We develop a pipeline for ESG news extraction and a state-of-the-art transformer based language model to predict stock volatility \cite{vaswani2017attention, devlin2019bert, lakshminarayanan2017simple}.
This pipeline can be generalized to other predicting targets with ease.
As a measure of price fluctuations and market risk, volatility plays an important role in trading strategies, investment decisions and position scaling.
In this paper, we focus on predicting Equity realized volatility, which is empirically calculated by the variance of observed returns of an asset. Volatility predictions often rely on predictive models based purely on price/return time-series, from standard statistical models of the GARCH family up to more recent deep-learning model based predictions ~\cite{yangliu2019}.

\begin{figure}[!htbp]
\centering
\includegraphics[width=0.9\textwidth]{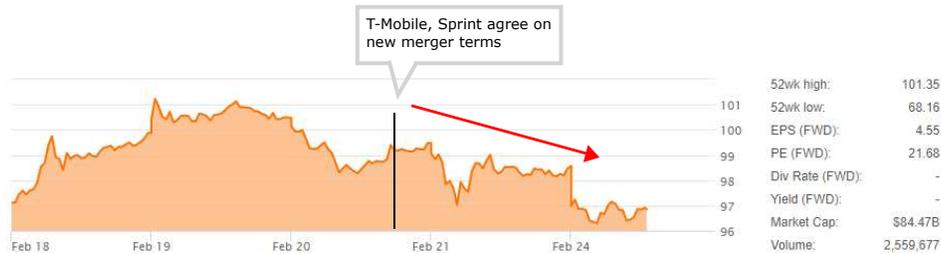}
\caption{An intuitive example of financial news and stock price movement from T-Mobile.}
\label{fig:news_stock}
\end{figure}

The input to our models is an alternative source of ESG information: textual financial news-flow.
Compared to structured ESG data provided by analysts or data vendors, ESG information from news-flow reflects more timely events of companies, and offers an alternative channel of capturing the relation of ESG events to market dynamics in a timely manner.
Numerous research demonstrated that financial news is closely related to market and is becoming a gold mine to analyze market participants' behaviour \cite{liu2018hierarchical, hu2018listening}. 
An intuitive example is illustrated in Fig.\ref{fig:news_stock}.

Though bearing rich information, ESG news is challenging to process for predicting models.
Raw textual data is categorical and symbolic represented, which is a hindrance for quantitative models.
Financial news is sparse in the sense that it moves in-parallel with real-world events in irregular timings.
This is in contrast to the structured and well-formatted market and factor data typically used in conventional quantitative models.
Although there are a variety of work studying predicting market behaviours with different data sources~\cite{guo2018bitcoin, weng2018predicting, schumaker2009textual, beck2019sensing}, how to exploit the predictive power of ESG news on volatility is rarely researched.

To this end, we resort to natural language processing (NLP) and deep learning techniques to explore the predictive power of ESG news. 
In particular, one key NLP technique, which helps hurdle the challenges above, is language representation (i.e. text embedding) \cite{vaswani2017attention, devlin2019bert, lan2019albert}.
Deep neural networks, e.g. recurrent neural networks and transformer, trained on large scale text corpus exhibit remarkable success in a variety of NLP applications, such as sentiment analysis, text matching, dialogue systems and so on.
This technique transforms text symbols into a numerically high-dimensional dense vectors, while importantly still preserving context and semantic relations.

\textbf{Contributions.}
Specifically, the contribution of this paper is as follows:
\begin{itemize}
    \item We propose a NLP and deep learning based pipeline for ESG news integration.
    \item We exploit the Transformer based language model to transform textual news into numerical representations including sentiments and semantic-preserving text embedding.
    \item The predicting model is based on Bayesian inference, in order to enable stable and robust predicting. 
    \item Evaluation on real data and different markets demonstrates the superior predicting performance as well as the efficacy of the volatility prediction in stock selection.
\end{itemize}

\section{ESG2Risk Framework}

\subsection{Overview}
In this part, we give an overview of the proposed ESG2Risk framework and define the notations used throughout the paper.

\begin{figure}[!htbp]
\centering
\includegraphics[width=0.97\textwidth]{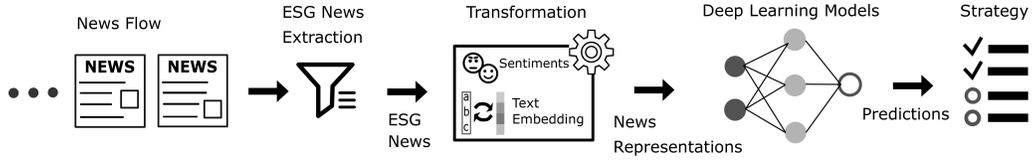}
\caption{Pipeline of the ESG2Risk Framework.}
\label{fig:framework}
\end{figure}

\textbf{Pipeline.}
Fig.~\ref{fig:framework} illustrates the pipeline mainly consisting of four components: ESG news extraction, transformation, deep learning models, and the strategy. The news flow represents streaming pieces of news, which can be obtained by a variety of tools, e.g. online news feeds, finance data vendors, web crawlers, and so on.

We first brief the main components of ESG2Risk as follows and then will mainly describe the components of ESG news extraction, transformation, and deep learning models in the next subsection.
\begin{itemize}
    \item ESG news extraction: 
    generic financial news flow rarely provides labelled ESG news and thus we developed the in-house extraction process.
    Streaming financial news first go through a filter, which makes use of an ESG vocabulary defined by our domain experts.
    The output is ESG related news and the corresponding companies mentioned in each piece of news.
    The mentioned companies are commonly provided as accompanied attributes by data vendors and can also be derived by linking entities in news to a stock dictionary.
    For instance, Fig.~\ref{fig:esg_topic} shows the top ESG topics in news within one week period.
    \begin{minipage}{\linewidth}
       \centering
       \includegraphics[width = 0.7\textwidth]{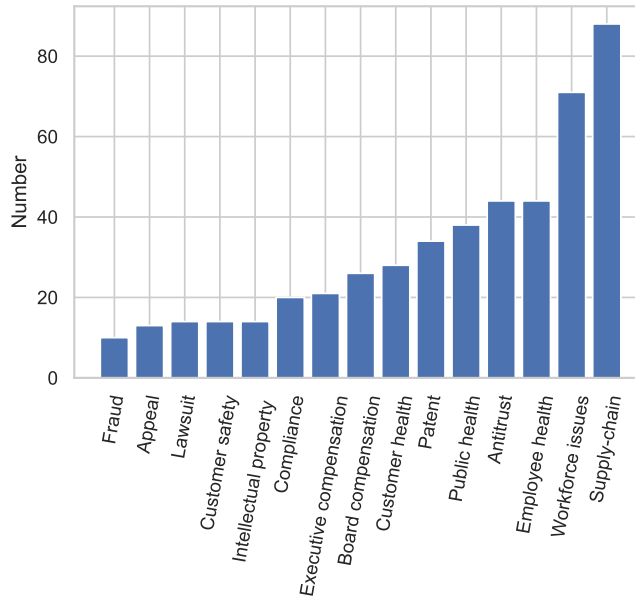}
       \captionof{figure}{Top-15 ESG topics in the news from an example time period April 06, 2020 to April 12, 2020.}
       \label{fig:esg_topic}
    \end{minipage}
    
    \item Transformation: 
    news is represented by symbolic textual data, which is unstructured and infeasible for quantitative models to process.
    As a results, this step aims to transform news into quantitative representations by transformer based language model and sentiment analysis \cite{vaswani2017attention, araci2019finbert, devlin2019bert}. 
    
    Language models or text embedding is a powerful technique in natural language processing (NLP), which is able to transform symbol represented text into numerical dense vectors preserving the semantic relatedness of text in the numerical space. 
    More details will be given in the following.
    
    \item  Models: 
    in the model training phase, we collect a dataset including pairs of numerical news representations and the volatility of the corresponding companies for supervised training of the predicting model. 
    
    In the inference phase of the production system, the newly arriving numerical news representations are fed into the trained models to predict the future volatility of the companies mentioned in the news.
    
    For both training and inference phases, we take advantages of Bayesian inference to enable stable learning and robust predicting \cite{lakshminarayanan2017simple, maddox2019simple}.
    Details will be presented in the next subsection.
    
    \item Strategy: 
    by using the volatility predictions from deep learning models, either stand-alone or combined with other signals, we design stock selection strategy.
\end{itemize}

\textbf{Problem Definition.} We formally define the problem of ESG news based volatility predicting as follows.

Market risk exists because of price changes~\cite{yangliu2019}.
The volatility of a stock $i$ is used to characterize the risk and the return fluctuation over time. 
Let $p_{i,t}$ be the price of stock $i$ at the end of a trading period $t$ with closing returns $r_{i,t}$ given by
\begin{equation}
r_{i,t} = \frac{p_{i,t}}{p_{i,t-1}} - 1 
\end{equation}
In this paper, we focus on the realized volatility, which is defined as:
\begin{align}
v_{i,t} = \sqrt{\frac{\sum_j r^2_{i, j}}{K_t}}
\end{align},
where $K_t$ is the number of return samples.

Define the universe of stocks as $\mathcal{I}$.
At time $t$, a set of stocks $I_t \subseteq \mathcal{I}$ is identified with ESG news in a sliding window w.r.t. $t$.
It means for a stock $i \in I_t$, it has ESG news mentions during time period $[t-w, t]$, where $w$ is the window length.
Then, the news mentioning $i$ at time $t$ is represented by $N_{i, t} = \{ n_{m} \}_{m \in M_{i, t}}$, where $M_{i, t}$ is the number of ESG news corresponding to stock $i$ in the time window $[t-w, t]$. $n_m$ denotes the text of one piece of news.

In the inference phase, given observed ${N_{i, t}}$, we predict the forward volatility $v_{i, t+\Delta}$ by a quantitative model $f(\cdot)$, namely $\hat{v}_{i, t+\Delta} = f(\text{Trm}(N_{i, t}))$, where $\text{Trm}(\cdot)$ is the operation of transforming unstructured news text into model-friendly numerical data and will be described in the next subsection. 
$\Delta$ is the forecasting horizon. 
Then, based on the set of predictions $\{\hat{v}_{i, t+\Delta}\}_{i \in {I}_t}$, we can further develop stock selection strategies.

In the learning phase, we collect paired news representations and ground-truth forward volatility, denoted by a dataset $\mathcal{D} = \{(N_{i, t}\, , v_{i, t+\Delta})\}_{i\in \mathcal{I}, t\in \mathcal{T}}$, to train the predicting model $f(\cdot)$ in the supervised way.

\subsection{News Representations}

In this part, we describe how to obtain numerical representations of news.
It is mainly based on pre-trained language models \cite{vaswani2017attention, devlin2019bert}.

Conceptually, the language model is designed to learn the usage of various words, phrases and how the language is written in general.
Technically, the typical building blocks in contemporary language models are recurrent neural networks, convolutional neural networks, or the more recently proposed Transformer architecture.
A variety of language models are developed by stacking certain types of building blocks with specialized training procedures.

During the training phase of a language model, each token of a text segment is initialized as a numerical dense vector.
These numerical vectors are fed into (stacked) building blocks to capture contextual relationships among words. 
Subsequently, the output vectors of building blocks are used in some language predicting tasks as the training objective, for instance, next word prediction, masked word prediction, next sentence prediction, etc.

\begin{figure}[!htbp]
\centering
\includegraphics[width=0.98\textwidth]{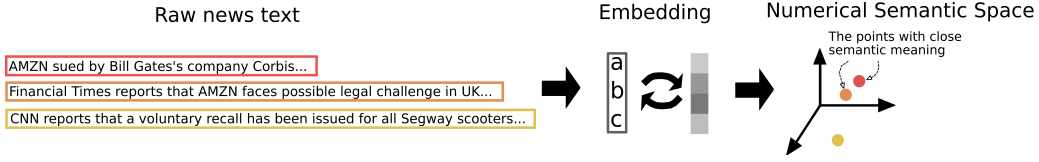}
\caption{An illustrative example of embedding text into numerical semantic space. 
Raw news text is transformed into dense vectors in the numerical space. 
The different colours represent the correspondence between text and vectors (i.e. points) in the space. 
Texts with semantic closeness are mapped to points close in the semantic space.}
\label{fig:embedding}
\end{figure}

A language model is typically pre-trained on a large corpus of text, for instance, the complete Wikipedia dump(2,500 million words), Book Corpus (800 million words), etc.
The training process aims to maximize the accuracy of these language predicting tasks, so as to learn vectors representing the semantic relations in large amount of text.
This pre-trained language model serves as the Swiss Army Knife for downstream NLP tasks. 
For instance, sentiment classification is proven to benefit from text embedding by pre-trained language models \cite{araci2019finbert}.
Refer to Fig.~\ref{fig:embedding} for an illustration of text embedding. 

In our case, given news $N_{i, t} = \{ n_{m} \}_{m \in M_{i, t}}$, the transformation is defined as follows:
\begin{equation}
\text{Trm}(N_{i, t}) \coloneqq (\text{pool}_s(\{\textbf{s}_m\}), \text{pool}_e(\{\textbf{e}_m\})),
\end{equation}
where $\mathbf{s}_m$ and $\mathbf{e}_m$ represent the sentiment information and text embedding of one piece of news. 

In particular, we use the sentiment analyzer trained on text embedding to extract sentiment scores over segments of news content, which form the fine-grained sentiment vector $\mathbf{s}_m$ for each piece of news.
By feeding ESG news text into a pre-trained Transformer based language model (e.g. BERT, RoBERTa, etc.)~\cite{devlin2019bert, lan2019albert}, the derived vectors $\{\mathbf{e}_m\}$ are quantitative-model-friendly and conveniently used in the subsequent learning of the volatility predicting model.

Note that different stocks have different number of ESG news over time.
We apply pooling operations to $\{\mathbf{s}_m\}$ and $\{\mathbf{e}_m\}$, in order to format them as structured data across stocks and time instants.
The derived dataset of the uniform shape will be ready for feeding the volatility model.
We choose the simple average pooling for $\text{pool}_s(\cdot)$ and $\text{pool}_e(\cdot)$, i.e. $\text{pool}(\{\textbf{x}_k\}) \coloneqq \frac{\sum \textbf{x}_k}{|\{ \textbf{x}_k \}|}$, while it is flexible to use different ones.

\subsection{Model Learning and Inference}

In this part, we present the architecture of the volatility model and Bayesian learning and inference. 

\begin{figure}[!htbp]
\centering
\includegraphics[width=0.63\textwidth]{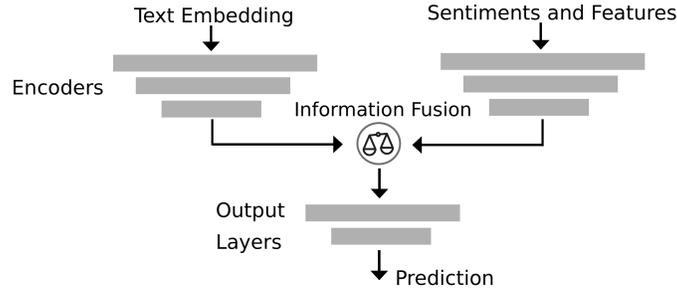}
\caption{Architecture of the Volatility Model Parameterized by $\Theta$.}
\label{fig:archi}
\end{figure}

Since we have two types of data, sentiment and text embedding from $\text{Trm}(N_{i, t})$, we design the encoder and information fusion architecture, as is shown in Fig.~\ref{fig:archi}. 
The idea is to capture data modal specific information by individual encoders and then to perform prediction using fused information from encoders.
Different information fusion strategies, e.g. attention mechanism, mixture, etc. can be applied~\cite{guo2019exploring}.
For encoders, we choose stacked dense layers (with residual connections)~\cite{iyyer2015deep}, though alternative encoders are free to choose.  

Instead of using the conventional squared error loss as the learning objective, we aim to learn the posterior of the model parameter in the Bayesian setting, as is defined below:
\begin{equation}\label{eq:posterior}
p(\Theta | \mathcal{D}) \propto \underbrace{\prod_{i, t} p_{\Theta}( v_{i, t+\Delta} | \text{Trm}(N_{i, t}) )}_{\text{Likelihood}} \cdot \underbrace{p({\Theta}}_{\text{Prior}}),
\end{equation}
where the set of parameters of the model to learn from data is denoted by the random variables $\Theta$.

In our task, the target variable volatility is essentially noisy and volatile.
More complexly, the sentiment and embedding vectors are high-dimensional, for instance, the embedding derived by BERT is a $768$-dimensional vector. 
These present a great challenge for conventional stochastic gradient descent based training algorithms to learn stable patterns in the data.

On the contrary, Bayesian style learning is probabilistic and theoretically designed to handle the stochasticity inherent in the data \cite{lakshminarayanan2017simple, maddox2019simple}.
Meanwhile, learning the posterior leads to ensemble inference, which aggregates predictions from a set of model realizations. 
It has been shown in many applications ensemble inference gives rise to more robust and accurate predicting performance~\cite{li2016preconditioned, lakshminarayanan2017simple, maddox2019simple}.
From the perspective of maximum a posterior probability (MAP) estimate, the prior in Eq.~\ref{eq:posterior} also functions as a regularization term to ensure the generalization ability of the learned model.

\begin{figure}[htbp]
\centering
\includegraphics[width=0.93\textwidth]{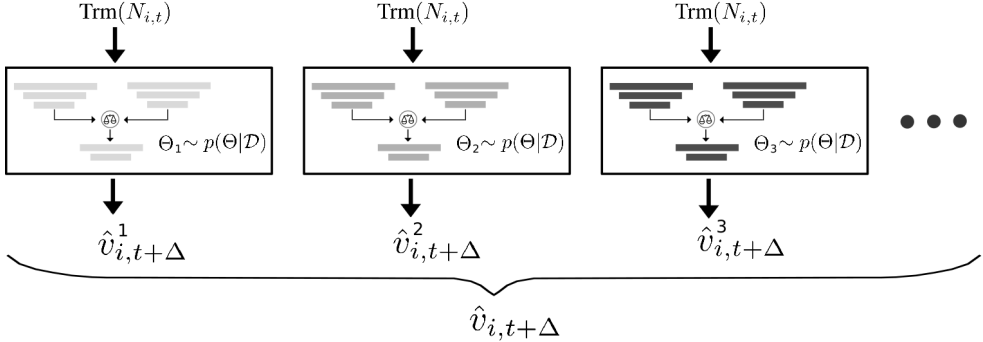}
\caption{Illustration of Bayesian Model Ensemble. 
The model in each block represents a realization of the model in Fig.~\ref{fig:archi} by the parameter sample $\Theta$.}
\label{fig:ensemble}
\end{figure}

Concretely, we use the modern stochastic gradient based Markov Chain Monte Carlo (SG-MCMC) method to obtain the approximate posterior of $p(\Theta | \mathcal{D})$ in the training phase \cite{li2016preconditioned, lakshminarayanan2017simple}.
Then, in the inference phase, given a testing set of news for stock $j$, the predictive probabilistic density of the volatility is derived as:
\begin{align}
\begin{split}
p(v_{j, t+\Delta} | \, \text{Trm}(N_{j, t}), \mathcal{D}) = \int_{\Theta} p_{\Theta}(v_{j, t+\Delta} | \, \text{Trm}(N_{j, t})) \cdot p(\Theta | \mathcal{D}) d\Theta \\
\end{split}
\label{eq:density_mixture}
\end{align}

Empirically, we use Monte Carlo method to sample parameter $\Theta_c$ from the posterior, thereby getting a set of model realizations, as is shown in Fig.~\ref{fig:ensemble}.
Each model realization provides the predictive mean of the volatility and the overall prediction is derived as:
\begin{equation}
\hat{v}_{j, t+\Delta} = \frac{1}{C} \sum_{c=1}^C \mathbb{E}[v_{j, t+\Delta} | \, \text{Trm}(N_{j, t}), \Theta_c] , \text{  } \Theta_c \sim p(\Theta | \mathcal{D})
\end{equation}

\section{Evaluation}

In this part, we report evaluation results to demonstrate the efficacy of our predicting pipeline. 

\textbf{Data}. 
From a collection of financial news in the time period from 2003 to 2019, we extract around 50 thousands ESG related news.
We link ESG news to companies from two different markets, i.e. MSCI US and All Cap EU. 
Then, for each market, we build the training and validation data using the time period from 2003 to 2014, while the rest is used as the out-of-sample testing data.
We evaluate the predicting performance of each market independently. 
The results reported below are from the testing data.
\setlength\extrarowheight{2pt}
\begin{table}[h]
  \centering
  \caption{One Week Forward Volatility Predicting Errors. \\
  Bold values indicate better performance.}
  \begin{tabular}{|c|c|c|c|c|}
    \hline
    \multirow{2}{*}{Market} & \multicolumn{2}{|c|}{RMSE} & \multicolumn{2}{|c|}{MAE} \\
    \cline{2-5}
        & Senti & ESG2Risk & Senti & ESG2Risk\\
    \hline
    MSCI-US  & 0.663 & \textbf{0.289} & 0.542 & \textbf{0.249} \\
    AC-EU  & 0.630 & \textbf{0.281} & 0.515  & \textbf{0.241} \\
    \hline
\end{tabular}
\label{tab:1w_error}
\end{table}

\setlength\extrarowheight{2pt}
\begin{table}[htbp]
  \centering
  \caption{Two Week Forward Volatility Predicting Errors.\\
  Bold values indicate better performance.}
  \begin{tabular}{|c|c|c|c|c|}
    \hline
    \multirow{2}{*}{Market} & \multicolumn{2}{|c|}{RMSE} & \multicolumn{2}{|c|}{MAE} \\
    \cline{2-5}
        & Senti & ESG2Risk & Senti & ESG2Risk\\
    \hline
    MSCI-US & 0.669 & \textbf{0.289} & 0.548 & \textbf{0.251} \\
    AC-EU   & 0.624 & \textbf{0.280} & 0.511  & \textbf{0.240} \\
    \hline
  \end{tabular}
  \label{tab:2w_error}
\end{table}

\textbf{Performance}.
Table~\ref{tab:1w_error} and ~\ref{tab:2w_error} exhibit the errors on two different predicting horizons, 1 week and 2 weeks. 
We report the rooted mean squared error (RMSE) and mean absolute error (MAE).
For comparison, the performance of the model using solely sentiments, i.e. Senti in the table, and our ESG2Risk, using both sentiment and text embedding, are reported.
It is observed that on both markets, ESG2Risk significantly outperforms the Senti method.

To assess the possibility to use our ESG2Risk predictions to build more attractive risk-adjusted return portfolios, we split the stocks based on quintiles of volatility predictions and calculate the average return of each quintile with 1 week and 2 weeks holding periods. 

\begin{figure}[!htbp]
\centering
\includegraphics[width=0.75\textwidth]{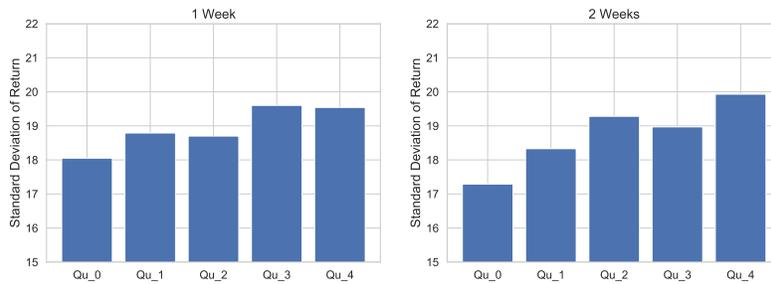}
\caption{MSCI US Quintile Portfolio Return Standard Deviation based on 1 and 2-week forward volatility predictions.}
\label{fig:std_us}
\end{figure}

\begin{figure}[!htbp]
\centering
\includegraphics[width=0.75\textwidth]{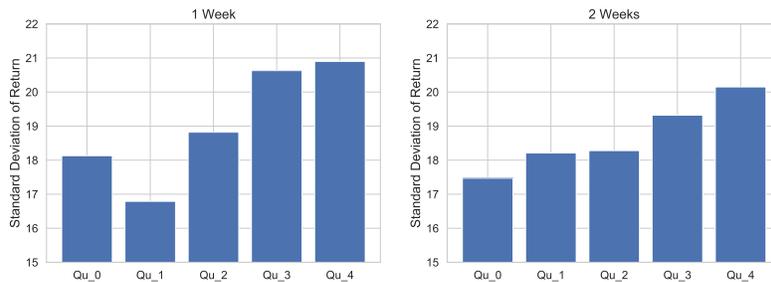}
\caption{All Cap EU Quintile Portfolio Return Standard Deviation based on 1 and 2 week forward volatility predictions.}
\label{fig:std_eu}
\end{figure}

\begin{figure}[!htbp]
\centering
\includegraphics[width=0.75\textwidth]{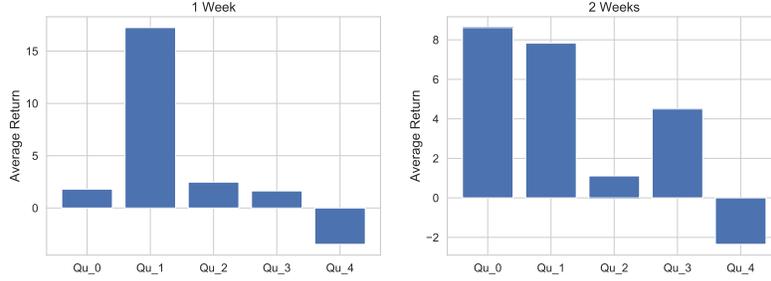}
\caption{MSCI US Quintile Portfolio Return based on 1 and 2-week forward volatility predictions.}
\label{fig:return_us}
\end{figure}

\begin{figure}[!htbp]
\centering
\includegraphics[width=0.75\textwidth]{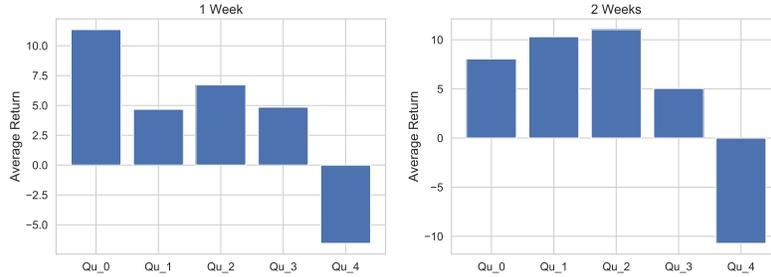}
\caption{All Cap EU Quintile Portfolio Return based on 1 and 2-week forward volatility predictions.}
\label{fig:return_eu}
\end{figure}

The average realized standard deviation of returns and returns of these quintile portfolios are shown in Fig.~\ref{fig:std_us} and \ref{fig:std_eu} and Fig.~\ref{fig:return_us} and \ref{fig:return_eu}.


 Quintile portfolios of stocks with high predicted volatility have significantly higher volatility than low predicted volatility portfolios in our out-of-sample test. The portfolio built with the highest predicted risk names (Qt 4 portfolio) also exhibits significantly lower returns, both in the MSCI US and All Cap Europe investment universes than the other quintile portfolios as shown in Fig.~\ref{fig:return_us} and \ref{fig:return_eu}. Extending the findings in existing work on structured ESG rating data ~\cite{de2015benefits}, integration of ESG news flow data can positively contribute to Equity portfolio returns.


\section{Conclusion}
In this paper, we implement a novel deep learning framework, ESG2Risk, to predict future volatility of stock prices. 
We show that a transformer-based language model successfully manages to extract information from ESG newsflow to predict future volatility of stock returns. Predictions of volatility in our model is more accurate when attempting to identify the stocks with the highest volatility risk in the market, hence the worst potential risk contributors to an Equity selection. 
Our research gives evidence that ESG newsflow does significantly impact future return and risk of companies and is a relevant factor for investors to consider when investing. Our findings in different geographies confirm that ESG newsflow integration can contribute to build profitable investment strategies, on top of improving the ESG profile of an Equity selection.

\bibliographystyle{plain}
\bibliography{reference} 

\begin{thebibliography}{10}

\bibitem{araci2019finbert}
Dogu Araci.
\newblock Finbert: Financial sentiment analysis with pre-trained language
  models.
\newblock {\em arXiv preprint arXiv:1908.10063}, 2019.

\bibitem{beck2019sensing}
Johannes Beck, Roberta Huang, David Lindner, Tian Guo, Zhang Ce, Dirk Helbing,
  and Nino Antulov-Fantulin.
\newblock Sensing social media signals for cryptocurrency news.
\newblock In {\em Companion Proceedings of The 2019 World Wide Web Conference},
  pages 1051--1054, 2019.

\bibitem{de2015benefits}
Indrani De and Michelle~R Clayman.
\newblock The benefits of socially responsible investing: An active manager’s
  perspective.
\newblock {\em The Journal of Investing}, 24(4):49--72, 2015.

\bibitem{devlin2019bert}
Jacob Devlin, Ming-Wei Chang, Kenton Lee, and Kristina Toutanova.
\newblock Bert: Pre-training of deep bidirectional transformers for language
  understanding.
\newblock In {\em Proceedings of the 2019 Conference of the North American
  Chapter of the Association for Computational Linguistics: Human Language
  Technologies, Volume 1 (Long and Short Papers)}, pages 4171--4186, 2019.

\bibitem{guo2018bitcoin}
Tian Guo, Albert Bifet, and Nino Antulov-Fantulin.
\newblock Bitcoin volatility forecasting with a glimpse into buy and sell
  orders.
\newblock In {\em 2018 IEEE International Conference on Data Mining (ICDM)},
  pages 989--994. IEEE, 2018.

\bibitem{guo2019exploring}
Tian Guo, Tao Lin, and Nino Antulov-Fantulin.
\newblock Exploring interpretable lstm neural networks over multi-variable
  data.
\newblock In {\em International Conference on Machine Learning (ICML)}, pages
  2494--2504, 2019.

\bibitem{hu2018listening}
Ziniu Hu, Weiqing Liu, Jiang Bian, Xuanzhe Liu, and Tie-Yan Liu.
\newblock Listening to chaotic whispers: A deep learning framework for
  news-oriented stock trend prediction.
\newblock In {\em Proceedings of the eleventh ACM international conference on
  web search and data mining}, pages 261--269, 2018.

\bibitem{iyyer2015deep}
Mohit Iyyer, Varun Manjunatha, Jordan Boyd-Graber, and Hal Daum{\'e}~III.
\newblock Deep unordered composition rivals syntactic methods for text
  classification.
\newblock In {\em Proceedings of the 53rd Annual Meeting of the Association for
  Computational Linguistics and the 7th International Joint Conference on
  Natural Language Processing (Volume 1: Long Papers)}, pages 1681--1691, 2015.

\bibitem{lakshminarayanan2017simple}
Balaji Lakshminarayanan, Alexander Pritzel, and Charles Blundell.
\newblock Simple and scalable predictive uncertainty estimation using deep
  ensembles.
\newblock In {\em Advances in neural information processing systems}, pages
  6402--6413, 2017.

\bibitem{lan2019albert}
Zhenzhong Lan, Mingda Chen, Sebastian Goodman, Kevin Gimpel, Piyush Sharma, and
  Radu Soricut.
\newblock Albert: A lite bert for self-supervised learning of language
  representations.
\newblock In {\em International Conference on Learning Representations}, 2019.

\bibitem{li2016preconditioned}
Chunyuan Li, Changyou Chen, David Carlson, and Lawrence Carin.
\newblock Preconditioned stochastic gradient langevin dynamics for deep neural
  networks.
\newblock In {\em Thirtieth AAAI Conference on Artificial Intelligence}, 2016.

\bibitem{liu2018hierarchical}
Qikai Liu, Xiang Cheng, Sen Su, and Shuguang Zhu.
\newblock Hierarchical complementary attention network for predicting stock
  price movements with news.
\newblock In {\em Proceedings of the 27th ACM International Conference on
  Information and Knowledge Management}, pages 1603--1606, 2018.

\bibitem{yangliu2019}
Yang Liu.
\newblock Novel volatility forecasting using deep learning–long short term
  memory recurrent neural networks.
\newblock {\em Expert Systems with Applications}, 132:99--109, 2019.

\bibitem{maddox2019simple}
Wesley~J Maddox, Pavel Izmailov, Timur Garipov, Dmitry~P Vetrov, and
  Andrew~Gordon Wilson.
\newblock A simple baseline for bayesian uncertainty in deep learning.
\newblock In {\em Advances in Neural Information Processing Systems}, pages
  13132--13143, 2019.

\bibitem{nagy2016can}
Zolt{\'a}n Nagy, Altaf Kassam, and Linda-Eling Lee.
\newblock Can esg add alpha? an analysis of esg tilt and momentum strategies.
\newblock {\em The Journal of Investing}, 25(2):113--124, 2016.

\bibitem{sassen2016impact}
Remmer Sassen, Anne-Kathrin Hinze, and Inga Hardeck.
\newblock Impact of esg factors on firm risk in europe.
\newblock {\em Journal of business economics}, 86(8):867--904, 2016.

\bibitem{schumaker2009textual}
Robert~P Schumaker and Hsinchun Chen.
\newblock Textual analysis of stock market prediction using breaking financial
  news: The azfin text system.
\newblock {\em ACM Transactions on Information Systems (TOIS)}, 27(2):1--19,
  2009.

\bibitem{vaswani2017attention}
Ashish Vaswani, Noam Shazeer, Niki Parmar, Jakob Uszkoreit, Llion Jones,
  Aidan~N Gomez, {\L}ukasz Kaiser, and Illia Polosukhin.
\newblock Attention is all you need.
\newblock In {\em Advances in neural information processing systems}, pages
  5998--6008, 2017.

\bibitem{weng2018predicting}
Bin Weng, Lin Lu, Xing Wang, Fadel~M Megahed, and Waldyn Martinez.
\newblock Predicting short-term stock prices using ensemble methods and online
  data sources.
\newblock {\em Expert Systems with Applications}, 112:258--273, 2018.

\end{thebibliography}

\newpage
Disclaimer

This document is not intended for persons who are citizens of, domiciled or resident in, or entities registered in a country or a jurisdiction in which its distribution, publication, provision or use would violate current laws and regulations. 
This publication has been prepared for general guidance on matters of interest only, and does not constitute professional advice. You should not act upon the information contained in this publication without obtaining specific professional advice. No representation or warranty (express or implied) is given as to the accuracy or completeness of the information contained in this publication, and, to the extent permitted by law, RAM Active Investments SA (RAM) does not accept or assume any liability, responsibility or duty of care for any consequences of you or anyone else acting, or refraining to act, in reliance on the information contained in this publication or for any decision based on it.
Furthermore, the information, opinions and estimates in this document reflect an evaluation as of the date of initial publication and may be changed without notice. Past performance must not be considered an indicator or guarantee of future performance, and the addressees of this document are fully responsible for any investments they make. The content of this document is confidential and can only be read and/or used by its addressee. RAM is not liable for the use, transmission or exploitation of the content of this document. Therefore, any form of reproduction, copying, disclosure, modification and/or publication of the content is under the sole liability of the addressee of this document, and no liability whatsoever will be incurred by RAM. The addressee of this document agrees to comply with the applicable laws and regulations in the jurisdictions where they use the information reproduced in this document. This document is issued by RAM. This publication and its content may be cited provided that the source is indicated. All rights reserved. Copyright 2020.

\end{document}